\documentstyle{article}
\font\big=cmbx12
\font\Big=cmbx12 scaled 1440
\font\Large=cmr10 scaled\magstep1
\def\preprint{\vtop{
                   \hbox{UMS/HEP 95-002}
                   \hbox{submitted to NIM}
                   \hbox{9 October 1995}}}
\begin{document}
{\Large
\rightline{\preprint}
\vskip 8pt
\centerline {\Big A Simple Multiprocessor Management System}
\centerline {\Big for Event-Parallel Computing}
\medskip
\centerline
{Steve Bracker, Krishnaswamy Gounder, Kevin Hendrix, and Don Summers}
\vskip 2pt
\centerline {Department of Physics and Astronomy}
\vskip 2pt
\centerline {University of Mississippi--Oxford}
\vskip 2pt
\centerline {University, MS 38677 \ USA}
}
\vskip 2pt
\leftline{\big Abstract}

Offline software using TCP/IP sockets to distribute particle physics events
to multiple UNIX/RISC workstations is described.  A modular, building block
approach was taken, which allowed tailoring to solve specific tasks
efficiently and simply as they arose. The modest, initial cost was having
to learn about sockets for interprocess communication.
This multiprocessor management
software has been used to control the reconstruction of eight billion raw data
events from Fermilab Experiment E791.

\bigskip
\leftline{\big 1. \ The E791 Reconstruction Task}

Fermilab Experiment E791 accumulated a large dataset
(50 Terabytes, 20 billion events, 24{\thinspace}000 8mm Exabyte tapes) in
1991 and early 1992 [1,2].
As might be expected,
the reconstruction and analysis of this data challenged available
computing resources; event reconstruction alone required over 10{\thinspace}000
mips-years of processing power. For $2{1\over2}$ years, reconstruction
processing was underway at four different locations [3]. The three
largest sites used clusters or {\it farms} of commercial UNIX/RISC
workstations
connected together by thin-wire Ethernet. Within
each farm, many processors operated together; data management and system
control were exerted from a single point via multiprocessor management
software. Here we describe the multiprocessor management software
developed and used at the University of Mississippi [4]. The Mississippi
farm hardware is shown in Figs. 1 and 2.

Most of the large-scale computing needs encountered in particle physics --
reconstruction and analysis of recorded data and generation of simulated
data -- are event-oriented. Each event's data packet is extracted from an
input stream and processed in isolation from other events. The results
from each event's analysis are merged into an output stream. The
computing power required to process each data packet is significant
relative to the time needed to transport the data, even over datapaths of
modest throughput. Many other scientific computing problems
(e.g.~small-object recognition in astronomical images) conform to the same
model. In all such problems, it is trivial to divide the total computing
task over many processors; each processor is simply given a share of the
events to process independently of its peers.
\smallskip

Most reconstruction and simulation tasks are developed and tested as
single-processor programs. They can be outlined in the following manner:

\bigskip
\leftline{initialize data structures}
\leftline{initialize input and output data streams}
\leftline{read run startup information from input data stream}
\leftline{read and store run calibration constants}
\vskip 5pt
\leftline{for every event from the input stream...}
\leftline{\hskip4mm unpack and check the input data}
\leftline{\hskip4mm {\it PROCESS THE EVENT}\ \  (most of the work)}
\leftline{\hskip4mm for good events,}
\leftline{\hskip8mm pack the processed event into output stream}
\vskip 5pt
\leftline{produce a report characterizing the processing job}
\leftline{stop}
\eject

When the program is ``nearly working,'' it is then moved into the
multiprocessing environment for final testing and production. If it has
been written in a sensible fashion, with dataflow management tasks cleanly
separated from event processing tasks, then the transition from single
processor to multiprocessor system should be relatively painless. Dataflow
management is generally vested in a single {\it server process}, while the
event processing is performed by numerous {\it client processes} running
in many processors. The server grants the clients access to calibration
data, reads the event input stream, parcels out events to the many client
processors, gathers output data from the clients, and writes the output
stream. The server processor is typically well endowed with peripheral
devices (disks, tape drives), whereas the clients may be very simple (but
powerful) processors with no peripherals except a network connection.

\bigskip
\leftline{\big 2. \ The Multiprocessor Manager's Tasks}

In the first implementations of event-oriented multiprocessing for
particle physics [5,6], the multiprocessor management software had to
do a great deal of hard work. Clients frequently had no significant
operating system, primarily because memory was expensive. Executable code and
calibration data had to be formatted in the server and downloaded into
the clients word by word. Extraction of input events from records and
merging of output events into records often had to be done in the server
because of limited client buffer memory, posing the danger of a
processing bottleneck at the server. Reports had to be retrieved from the
clients as tables of numbers, to be formatted and output by the server.
Substantial effort was often required to move a single-processor program
onto a multiprocessor system, because the division between server code and
client code was intricate. Developing the multiprocessor management code
could become nearly as large a task as developing the event processing
software: an odious burden.

With the advent of cheaper memory (enabling each client to have a real
operating system and large event buffers) plus good networking
software, most of the difficulties inherent in moving code to a
multiprocessor vanished. Rather than relating ``war stories'' of heroic
efforts and brilliant strategies leading to victory in the face of
staggering difficulties, this paper celebrates the fact that
multiprocessor management is now straightforward, that an effort tiny
compared to the development of application code will suffice to
distribute the workload efficiently and reliably among many processors,
and that generic system software nearing completion promises to make the
task even easier in the future.

\bigskip
\leftline{\big 3. \ A Disk-Based Multiprocessor Manager}

It is easiest to understand the multiprocessor strategy by considering
first a disk-based system. The clients each have a real operating system
(in our case ULTRIX, a flavor of UNIX) and thus can read and write disk files,
but they have no disks attached directly to them. Instead, clients have
access over the network to disks attached to the server. Using Network
File System (NFS) software, disks or portions of disks can be
``cross-mounted'' so that they are accessible by multiple processors.
For us, this meant the server itself and all of its clients.
NFS is supplied as
standard software with many types of workstations, and is available as an
option on most of the rest.

The fact that clients, though diskless, have full access to disk services
across the network, immediately solves several problems that burdened
earlier systems:

\newcounter{bean1}
\begin{list}
{(\arabic{bean1})}{\usecounter{bean1}
\setlength{\rightmargin}{14mm}
\setlength{\leftmargin}{20mm}}

\item
Clients can read executable code and start programs; it is no longer
necessary for the server to micromanage the downloading and startup of
clients, though high level control of client processes is still maintained
in the server,

\item
Clients can read their own calibration files; it is no longer
necessary for the server to explicitly read, reformat, and transmit
calibration data for the clients,

\item
Clients can write their own report files; it is no longer necessary
for the server to explicitly extract, format, and write reports for the
clients.
\end{list}

\smallskip
In many cases, cross-mounted disks can even be used to provide for the movement
of event data (input and output) between server and clients:

\begin{list}
{(\arabic{bean1})}{\usecounter{bean1}
\setlength{\rightmargin}{14mm}
\setlength{\leftmargin}{20mm}
\setcounter{bean1}{3}}

\item
The server can read events from tape and write them to disk; the
clients can read the events from disk and process them. For example, each
client can own two input files; the server inserts events into one while the
client consumes events from the other,

\item
The clients can write output events to disk; the server can read the
disk files, and write the events to output tapes. Each client can
own two output files;
the client inserts processed events into one while the
server removes events from the other and writes them to tape.
\end{list}

Referring back to the outline of the processing tasks, it is clear that
the program running in each client of a multiprocessor system is very similar
to the
single-processor program. Each client loads its own code and starts it.
Each client reads its own calibration constants from disk. Each client
writes its own reports to disk. Each client reads its input data from
disk and writes its output data to disk, just as single-processor code
usually does during program development. The server's tasks now become
the following:

\newcounter{bean2}
\begin{list}
{(\arabic{bean2})}{\usecounter{bean2}
\setlength{\rightmargin}{14mm}
\setlength{\leftmargin}{20mm}}

\item
Keeping a list of the available processors and requesting them to start
instances of the processing program,

\item
Making startup data (e.g. run number) available to clients in a disk file,

\item
Assuring that valid calibration constants are available to be read by
the clients,

\item
Reading input data from tape and distributing events to the various
client disk files,

\item
Fetching output data from the various client disk files and writing
it to tape,

\item
Gathering reports from the client report files and producing (where
appropriate) a system-wide report.
\end{list}

This scheme was very simple to implement, maintain, and document. For
applications in which dataflow is slow, it works very well. Clients
are provided with certified events (or blocks of events) on a reliable
medium, protected from the vagaries of tape reading, which is handled by
the server. Likewise, clients have a reliable medium for writing output;
problems with tape writing are handled in the server. If enough disk is
available, whole tape files can be staged through the disk files, leading
to more reliable performance from streaming tape drives [7].

\bigskip
\leftline{\big 4. \ A TCP/IP Sockets Based Multiprocessor Manager}

As a farm's data throughput is increased by adding processors, upgrading
processors, or making application code run faster, passing the event stream
through disk can become a bottleneck; clients may spend too much time
waiting for input events to be read and output events to be written, and
disk ``thrashing'' (incessant head motion) may set in. One possible
solution is to move the event streams through cross-mounted {\it memory
disks} using special software by which a portion of the server's memory
is set aside and appears to the user as a very high-speed, low-latency,
non-thrashing, disk drive. When we explored this option in 1992, such
software was available but did not perform well; it is possible that
suitable products now exist. Memory disks preserve the advantage that the
application code for a multiprocessor system
looks almost identical to that for a
single processor; all data passing is done with normal Fortran reads and
writes.

Instead, our multiprocessor manager bypasses disk altogether for those
portions of the dataflow that are fast enough to challenge the disk's
throughput; data is moved between processes using TCP/IP
(Transmission Control Protocol/Internet Protocol) network
services [8].
Using this facility, processes can establish a connection
between themselves and pass data back and forth by {\it
read\_from\_connection} and {\it write\_through\_connection} subroutine
calls. Event input/output can no longer be implemented as simple Fortran
reads and writes in the client (a disadvantage), but on the other hand,
the high throughput of direct network data transfers becomes available.

Before deciding to implement a TCP/IP-based scheme, we had to answer
three significant questions:

\newcounter{bean3}
\begin{list}
{(\arabic{bean3})}{\usecounter{bean3}
\setlength{\rightmargin}{14mm}
\setlength{\leftmargin}{20mm}}

\item
Would passing data through TCP/IP yield a significant improvement in
throughput over NFS disk-based data transfers?

\item
Would resulting modifications to the client code be tractable?

\item
Given very limited time to produce the management code, would TCP/IP
be easy enough to learn and use, especially for those of us coming from a
VAX/VMS---Fortran environment?
\end{list}

We wrote a test package to shuttle data between several processors
to measure throughput and reliability. The results: average data
throughputs in excess of 900 kilobytes per second (almost full Ethernet
speed) could be maintained (we
needed only 150 kilobytes per second); the impact on client efficiency was
immeasurably low at maximum projected throughput; and data corruption was
not seen.

In keeping with the Fortran orientation of the experiment's software,
Fortran-callable functions were written in C for all of the system
services needed to support TCP/IP data transfers. As a result, the
changes to the client code were minimal and easily understandable to
Fortran-only programmers.

Fig.~3 illustrates how the network I/O calls are used. As the server
prepares to start a client, it uses {\it make\_socket} to ``have a phone
put in'', so that it will be able to connect to the client. When the
client starts, it too uses {\it make\_socket} to ``have a phone put in''.
The server ``lists its number'' by binding its socket to a port ({\it
bind\_socket}), and ``stays near the phone'' listening for an attempt to
connect ({\it listen\_socket}). The client ``calls up'' the server ({\it
connect\_socket}) and the server ``picks up the phone'' establishing the
connection ({\it accept\_socket}).

When it needs input data, the client ``places its order'' by writing a
message to the server ({\it write\_socket}). The server is continually
monitoring all of the client connections for requests ({\it
select\_socket}). When a request comes in, the server ``writes down the
order'' ({\it read\_socket}), and does its best to satisfy the client's
request. The client and server shuttle messages back and forth (each
writing to the other and reading from the other) until the input data is
exhausted. At that point, the server notifies the clients that there is
no more data. The clients then finish their tasks, close their connections
({\it close\_socket}), and exit; the server finishes its tasks and exits.

The Fortran-callable routines that manipulate sockets and connections
really are that simple to use. Only one routine ({\it read\_socket}) is
any more than a C to Fortran interface, and even {\it read\_socket} is
trivial. Almost all of the real work involved with network communications
has already been done in UNIX, TCP/IP, and Berkeley Sockets.

\bigskip
\leftline{\big 5. \ Performance}

Only the transport of input events from server to clients was implemented
in TCP/IP; output events from clients to server needed to be staged to disk
to make best use of our Exabyte tape drives,
so they were written to
disk through NFS as before. This output scheme was efficient because five out
of six events were rejected by a filter after reconstruction and were not
output.

Tape reading was multiply buffered, so that events were almost always
available immediately when a client requested them. We used a simple trick
to ensure that clients were not suffering from delays in receiving new
events. Whenever input data was available, the server checked the clients to
see which ones were asking for new input. The server always checked the
clients in the same order, so processors at the start of the list had
priority over processors near the end of the list. If clients were
``spinning,'' i.e.~waiting for events, this was reflected in anomalously low
throughput in the least favored processors. By this and several
additional measures, it appears that more than 97\% of the client cycles
were being put to beneficial use (actually processing events)
whenever the farm was running at all.

Funding awarded in June 1993 allowed an expansion of the Mississippi computing
facility from 1100 to 2900 mips. By July 1993, the increased computing power
had been acquired and was processing data.
E791 reconstruction was completed in September 1994.
A total of eight billion events on 10{\thinspace}000 raw data tapes were
processed
at the University of Mississippi.
Preliminary versions of the reconstruction software were run to find out
which E791 algorithms would yield the most physics per tape. One result was
a tripling in the yield of charm particles.  When the
final reconstruction software was ready, it was run.
Overall efficiency,
considering all cycles lost for {\it any} reason, exceeded 90\%.
%The processors were divided into four farms of roughly equal computing power;
%each farm ran a separate job stream. The tape changing requirements were
%matched to the local mode of operation.
%Dual inputs (two tape drives which alternate) and dual outputs (one
%scratch disk drive and one tape drive) were employed to avoid
%continuous supervision by an operator.
%This dual input and output system led to fewer lost
%CPU cycles from mechanical failures,
%than automatic tape changers, which were used initially.
Thus the multiprocessor
manager has proven itself to be efficient and robust under varying
conditions in what we believe to
be a fairly typical university operating environment.

For the E791 experiment as a whole, the successful management of multiple
processors has provided the full reconstruction of over 200{\thinspace}000
particles with a charm quark. This large charm sample is in turn generating
new physics results [9--15].

\bigskip
\leftline{\big 6. \ Scavenging Computing Cycles}

Although most of the processors in the Mississippi system are mounted in
racks at a central location, some are at people's desks
and serve as
general purpose workstations. Many workstation activities -- editing,
compiling and running small programs, reading and writing e-mail, etc. --
can coexist with the farm client process running in the background.
However, there are some workstation activities that are incompatible with
farm operations, and the client process must be removed from the
workstation. Nevertheless, it is certainly helpful to be able to
scavenge the desk workstation cycles when they are otherwise unused.

There are several approaches to using the desk workstations as farm
clients. On one extreme, the workstation can be removed from the farm
client list; it will never be used as a farm client. At the other
extreme, one can forbid workstation activities that interfere with
the farm operation. In between there is ample middle ground. For example,
it is possible to have the farm server examine the workstation activities
from time to time and adjust the priority or run status of the farm
client process accordingly.

At Mississippi, we have found it satisfactory to allow users to abruptly
kill the client process whenever they find its activities on their
workstations to be troublesome. The server is quickly aware that the
client has disappeared and adjusts its event distribution accordingly. If
a workstation user knows that he will be engaged in a computationally intensive
activity for a long time, he may edit the client list and remove his
workstation from the farm. Once killed, a workstation cannot participate
as a farm client again until the next job is started. Though crude, this
scheme has proven highly effective in our operating environment.

One disadvantage of this approach is that a few input events may be
``trapped'' in a killed processor, and thus lost. We take a rather
cavalier attitude toward such losses; with twenty billion events, we can
afford to lose a few now and again. Although in principle the server can
hold event buffers until successful processing is assured, and reassign
the events to another processor if they become trapped in a disabled
client, we don't do that. In E791, processing raw events is rather like
hauling corn to market in a truck. If a few grains of corn fall out of
the truck, no one becomes concerned until the loss becomes large enough
to be economically important. The alternative approach -- treating events
as babies in a hospital nursery, where one normally expects a somewhat
stricter accounting -- only makes sense when the events have become
greatly enriched in scientific significance, late in the reconstruction
and analysis cycle.

\bigskip
\leftline{\big 7. \ Multiprocessor Operations}

The most vexing operational problems are those that one might expect in
handling a dataset this large~-- ensuring that all tapes are processed
exactly once, preparing and maintaining run-dependent calibration files,
making sure that all of the output tapes are correctly labeled, preparing
and examining the necessary report files, etc. During the first year of
operation, job flow was controlled by scripts composed with the help of
small interactive programs.
In 1993, a farm job manager with an X Window
graphical user interface was written.
It was specifically targeted at preventing
errors we had observed to
occur in
the setup of jobs and
the management of tapes. Most of the day-to-day system operation in later
stages was
performed by students, many of whom had little understanding of the
internal workings of the system.

Farm management software consists of three independent programs:

\newcounter{bean4}
\begin{list}
{(\arabic{bean4})}{\usecounter{bean4}
\setlength{\rightmargin}{14mm}
\setlength{\leftmargin}{20mm}}

\item
The server code, which provides system and dataflow control,

\item
The client code, linked to the application code routines,

\item
The tape-writing code, running on the server, which fetches events
written to disk by clients and transfers them to tape.
\end{list}

The server, the tape writer, and each client produces report files for each
job. The tape writer (the last program to handle the data) gathers
these reports together and produces two farm-wide reports. One is a
standard statistics file common to all E791 computing sites; the other is
a report tailored to the needs of the Mississippi site.

\bigskip
\leftline{\big 8. \ Future Directions}

Event-parallel computing farms have come to dominate large-scale
computations in experimental particle physics. Because computing paradigms
seldom remain viable for more than a decade or so in this field, it is
perhaps useful to ask ``Whither computer farming?'' in the next few years.

Two trends in the computer industry make it unlikely that ``farming as
usual'' will continue much longer. First, individual workstations, PCs,
and Macintoshes
now coming to market offer such prodigious computing power that each processor
can exploit the full I/O bandwidth of its data storage peripherals; in
that case, there seems little point in concentrating dataflow through a
server, although one might still imagine a single locus of system control
for several computers, each directly attached to its own peripheral
devices [16].

Second, with the increased popularity of object-oriented design and
languages such as C++ which support it, the data structures now being
explicitly passed between processes will be implemented as objects of
classes.  Very soon it will be possible to define remote objects,
which will tie
together the resources of many processors within a single programming
environment. For programs written within such a paradigm,
there will be
almost no difference between a single processor implementation and a
multiprocessor implementation except for listing
the computing resources
that may be brought to bear on the task.

In the early days of ``computer farming'' in particle physics, there were
not suitable commercial processors available, so we built our own [5,6]. After
a short time, we were put out of the processor building business by
high-powered workstations offered by several vendors [17]. The early
implementations of multiprocessor management software were complex, costly,
and cumbersome to use; the need for them has been snuffed out by the
widespread availability of interprocess communication tools such as TCP/IP
sockets and NFS. The simple streamlined multiprocessor management
toolkits remain and are likely to be in use for a bit longer, but their
end is also in sight. For tasks with mammoth computational needs and
modest dataflow rates, truly transparent, vendor-independent,
object--oriented access to the
combined power of dozens or hundreds of inexpensive powerful processors
appears to be imminent; ``computer farm management'' as we now practice
it, is an idea whose time has come and nearly gone.

\bigskip
\leftline{\big Acknowledgements}

We especially thank Lucien Cremaldi and Breese Quinn for their contributions
to building and running the Mississippi farm.
This effort was partially supported by the United States Department of Energy,
DE-FG05-91ER40622.
\bigskip

\bigskip
\leftline{\big References}
\newcounter{bean5}
\begin{list}
{[\arabic{bean5}]}{\usecounter{bean5}}
\item
D.J.~Summers et al., {\it Charm Physics at Fermilab E791}, Proceedings of the
XXVII$^{\rm th}$ Recontre de Moriond, Electroweak Interactions and Unified
Theories, Les Arcs, France (15-22 March 1992) 417; presents an overview of the
experiment.
\item
S. Amato, J.R.T.~de Mello Neto, J.~de Miranda, C.~James, D.J.~Summers, and
S.B.~Bracker, {\it The E791 Parallel Architecture Data Acquisition System},
Nucl.~Instr.~and Meth. A324 (1993) 535;
describes the system used to collect the E791 dataset.
\item
The Ohio State University (moved to Kansas State University in 1993);
Fermi National Accelerator Laboratory; Centro Brasileiro de Pesquisas
F{\'i}sicas--Rio de Janeiro; the University of Mississippi.
\item
Although all four processing centers ran the same physics reconstruction code,
three different multiprocessor management systems were used.
\item
Paul F.~Kunz et al.,
{\it Experience Using the 168/E Microprocessor for Off-line Data Analysis},
IEEE Trans. Nucl. Sci. 27 (1980) 582.
\item
J.~Biel et al., {\it Software for the ACP Multiprocessor System},
Proc.~of the Intl.~Conf.~on Comp.~in High Energy Physics, Asilomar
(2-6 Feb.~1987), Comp.~Phys.~Comm.~45 (1987) 331; FERMILAB-Conf-87/22.
\item
Exabyte tape drives operate most reliably if they read and write data
continuously rather than starting and stopping incessantly. In addition, head
wear depends on the time the tape is tensioned over the rotating head, not the
amount of data written. You obtain far more useful life (Terabytes per head
change) from an Exabyte that reads and writes at nearly full speed.
\item
Sidnie Feit, {\it TCP/IP: Architecture, Protocols, and Implementation},
McGraw-Hill, Inc. (1993), ISBN \hbox{0-07-020346-6,} Chapter 17.
There is also an extensive bibliography.
\item
A.K.S.~Santha, {\it Study of the Decay
$D^0 \rightarrow K^0_{\rm{S}} K^{\pm} \pi^{\mp}$},
APS/AAPT Joint Meeting, Crystal City, VA (18--22 April 1994);
\quad Bull.~Amer.~Phys.~Soc. 39 (1994) 1030.
\item
Tom Carter, {\it Production Asymmetries in $x_f$ and $P_t^2$ for $D^\pm$
Mesons}, Proc.~of the 8th Meeting, Div.~of Particles and Fields of
the Amer.~Phys.~Soc., Albuquerque, NM (2--6 August 1994) 513.
\item
Milind Purohit and James Wiener, {\it Preliminary Results on the Decays
$D^+ \rightarrow K^+ \pi^+ \pi^-$,  \ $D^+ \rightarrow K^+ K^+ K^-$},
Proc.~of the 8th Meeting, Div.~of Particles and Fields of the Amer.~Phys.~Soc.,
Albuquerque, NM (2--6 August 1994) 969.
\item
Guy Blaylock, {\it A Search for $D^0 \overline{D}{\thinspace}^0$
Mixing at the Fermilab E791 Experiment},
Proc.~of the 8th Meeting, Div.~of Particles and Fields of
the Amer.~Phys.~Soc., Albuquerque, NM (2--6 August 1994) 980.
\item
Krishnaswamy Gounder and Lucien Cremaldi, {\it D-Meson--Pion Production
Correlations}, APS/AAPT Joint Meeting, Washington, DC (18--21 April 1995);
\quad Bull. Amer. Phys. Soc. 40 (1995) 1022.
\item
Arun Tripathi, {\it Search for $D^0$--$\overline{D^0}$ Mixing in Semileptonic
Decays}, APS/AAPT Joint Meeting, Washington, DC (18--21 April 1995);
\quad Bull.~Amer.~Phys.~Soc. 40 (1995) 1035.
\item
E.M.~Aitala et al., {\it Search for the Flavor-Changing Neutral Current Decays
$D^+ \rightarrow \pi^+ \mu^+ \mu^-$ and $D^+ \rightarrow \pi^+ e^+ e^-$},
KSU HEP-95-01, FERMILAB Pub-95/142-E (submitted to Phys.~Rev.~Lett.).
\item
One caveat. I/O is getting faster but for now not as quickly as CPUs.  We used
Ethernet at 10 Mbit/s; and Exabyte 8200 and 8205 tape drives storing 2
Gigabytes at 0.25 Mbytes/s. Full Duplex Fast Ethernet now runs at 2 $\times$
100 Mbits/s. The Quantum DLT4000 tape drive now stores 20 Gigabytes at 1.5
Mbytes/s.  The Exabyte Mammoth Tape drive ({\it{still in development}}) is
designed to store 20 Gigabytes at 3 Mbytes/s.  The current Exabyte 8505XL
stores 7 Gigabytes at 0.5 Mbytes/s.  Unhappily, the cost of tape per Terabyte
stored has not decreased.
\item
C.~Stoughton and D.J.~Summers, {\it Using Multiple RISC CPUs in
Parallel to Study Charm Quarks}, Computers in Physics 6 (1992) 371.
\end{list}
\vfill
\eject

\null
\vskip195mm
\includegraphics{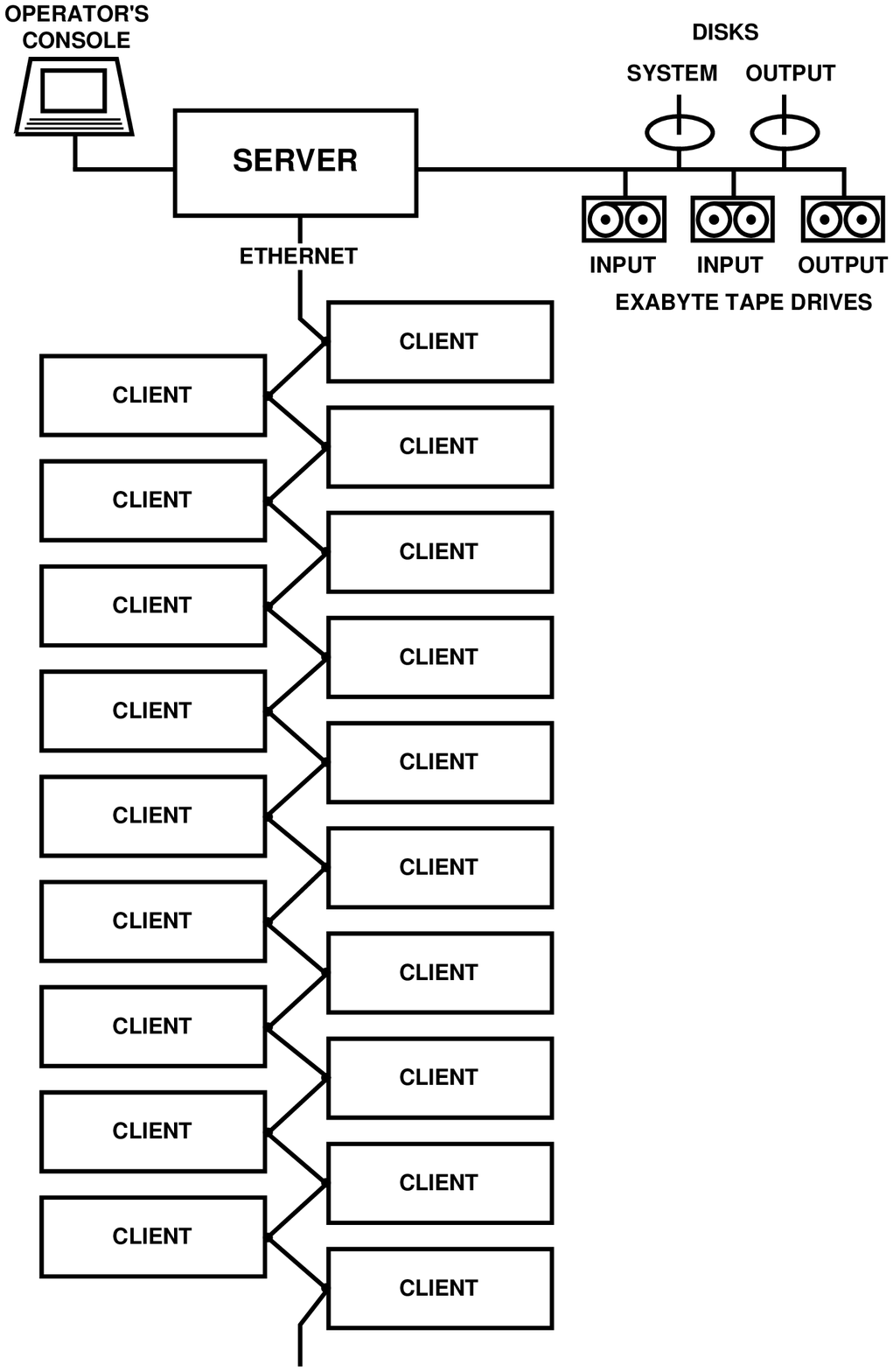}
\vfill
{\vbox{\parindent=0pt
Fig.~1. Computing farm configuration at the University of Mississippi.
Servers and clients are DECstation 5000 workstations running ULTRIX.
Some have MIPS R3000 Processors; others have the more powerful
MIPS R4000. Altogether, there are 68 processors organized into
four farms, each with a separate job stream.  One typical farm is shown in
this diagram.
The two input tape drives alternate.  The output is staged through disk to
tape.  This I/O scheme avoids the need for continuous operator supervision.
The total computing power is about 3000 mips.
}}
\eject

\null
\vskip195mm
%\special{ psfile=figure2.eps hscale=102 vscale=102 voffset=-310  hoffset=-80}
\vfill
{\vbox{\parindent=0pt
Fig.~2. A photographic overview of the University of Mississippi computing
farm.  Servers are on the four tables.  Clients are on the racks shown as
well as on desktops which are not shown.  The espresso machine is on the left.
}}
\eject

\null
\vskip216mm
\includegraphics{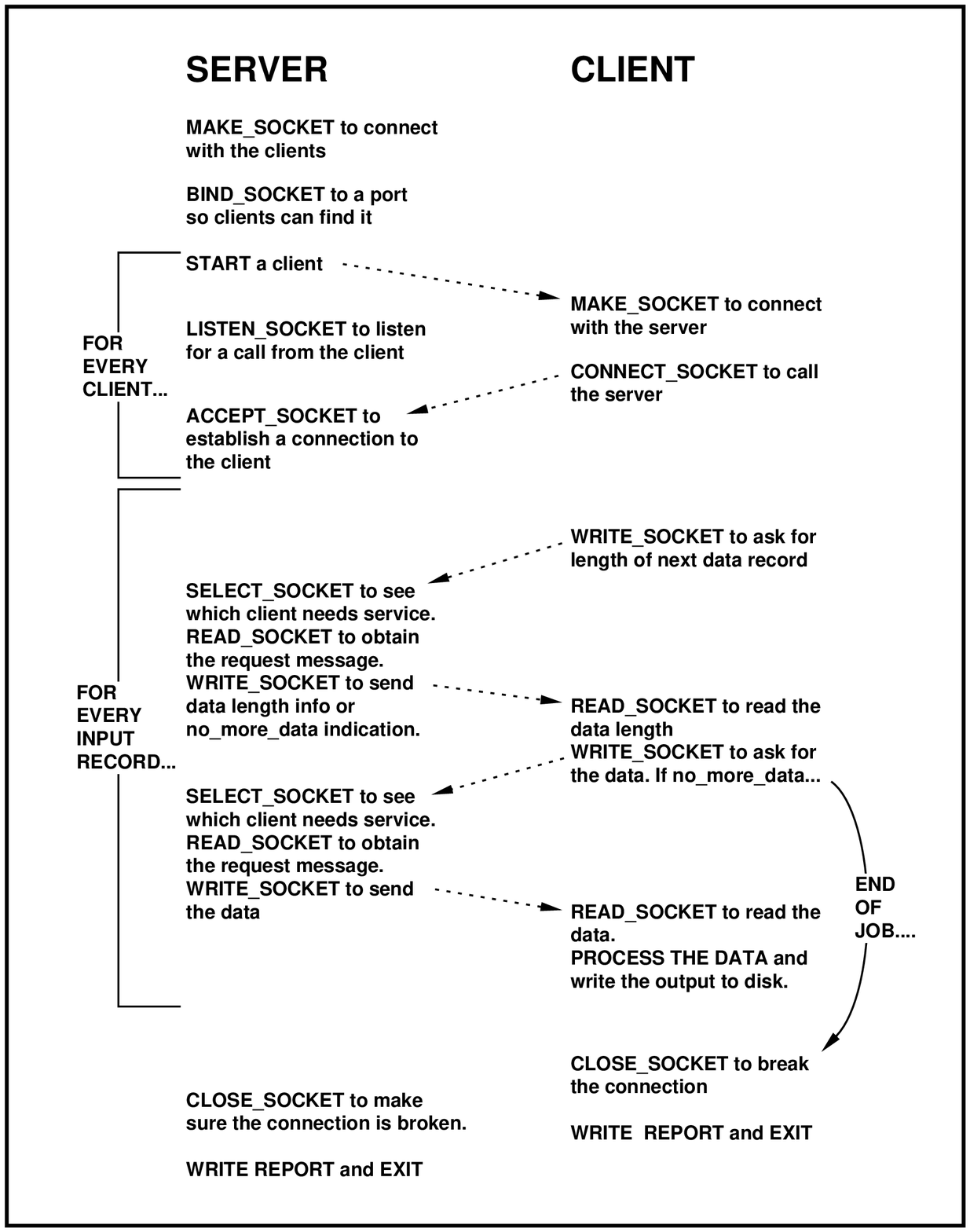}
\vfill
\leftline{Fig.~3. Communication between server and clients using TCP/IP
sockets.}

\end{document}